\documentclass[apj]{emulateapj}

\begin{document}

\submitted{Accepted to ApJ, December 14, 2013}

\title{A \textit{Spitzer} Search for Transits of Radial Velocity Detected Super-Earths}

\author{
J.~A.~Kammer\altaffilmark{1,*}, 
H.~A.~Knutson\altaffilmark{1},
A.~W.~Howard\altaffilmark{2},
G.~P.~Laughlin\altaffilmark{3},
D.~Deming\altaffilmark{4},
K.~O.~Todorov\altaffilmark{5}, 
J.-M.~Desert\altaffilmark{6,1,11},
E.~Agol\altaffilmark{7},
A.~Burrows\altaffilmark{8}, 
J.~J.~Fortney\altaffilmark{3}, 
A.~P.~Showman\altaffilmark{9}, 
N.~K.~Lewis\altaffilmark{10,11}
}

\altaffiltext{1}{Division of Geological and Planetary Sciences, California Institute of Technology, Pasadena, CA 91125, USA}
\altaffiltext{2}{Institute for Astronomy, University of Hawaii, Honolulu, HI 96822, USA}
\altaffiltext{3}{Department of Astronomy and Astrophysics, University of California at Santa Cruz, Santa Cruz, CA 95064, USA}
\altaffiltext{4}{Department of Astronomy, University of Maryland at College Park, College Park, MD 20742, USA}
\altaffiltext{5}{Institute for Astronomy, ETH Z\"{u}rich, 8093 Z\"{u}rich, Switzerland}
\altaffiltext{6}{CASA, Department of Astrophysical and Planetary Sciences, University of Colorado, Boulder, CO 80309, USA}
\altaffiltext{7}{Department of Astronomy, University of Washington, Seattle, WA 98195, USA}
\altaffiltext{8}{Department of Astrophysical Sciences, Princeton University, Princeton, NJ 08544, USA}
\altaffiltext{9}{Lunar and Planetary Laboratory, University of Arizona, Tucson, AZ 85721, USA}
\altaffiltext{10}{Department of Earth, Atmospheric, and Planetary Sciences, Massachusetts Institute of Technology, Cambridge, MA 02139, USA}
\altaffiltext{11}{Sagan Fellow}
\altaffiltext{*}{jkammer@caltech.edu}

\begin{abstract}
Unlike hot Jupiters or other gas giants, super-Earths are expected to have a wide
variety of compositions, ranging from terrestrial bodies like our
own to more gaseous planets like Neptune. Observations of transiting
systems, which allow us to directly measure planet masses and radii
and constrain atmospheric properties, are key to understanding the
compositional diversity of the planets in this mass range. Although
Kepler has discovered hundreds of transiting super-Earth candidates over the
past four years, the majority of these planets orbit stars that are
too far away and too faint to allow for detailed atmospheric characterization
and reliable mass estimates. Ground-based transit surveys focus on
much brighter stars, but most lack the sensitivity to detect planets
in this size range. One way to get around the difficulty of finding
these smaller planets in transit is to start by choosing targets that
are already known to host super-Earth sized bodies detected using
the radial velocity technique. Here we present results from a \textit{Spitzer} program to observe six of the most
favorable RV-detected super-Earth systems, including HD~1461, HD~7924,
HD~156668, HIP~57274, and GJ~876. We find no evidence for transits
in any of their 4.5 $\mu$m flux light curves, and place limits on the
allowed transit depths and corresponding planet radii that rule out
even the most dense and iron-rich compositions for these objects. We also
observed HD~97658, but the observation window was based on a possible ground-based
transit detection \citep{henry2011detection} that was later ruled out; thus
the window did not include the predicted time for the transit detection recently made by \textit{MOST}
\citep{dragomir2013detects}.

\end{abstract}


\section{Introduction}

Super-Earths are a unique class of planets that have masses ranging between that of Earth and Neptune. They may form via diverse pathways \citep[e.g.,][]{hansen2012migration,chiang2013minimum}, and current observational constraints indicate a wide range of bulk densities and compositions for these planets \citep{valencia2010composition,valencia2013bulk,fortney2013framework}. By characterizing the properties of these unique worlds, which have no solar system analogue, we can learn more about their physical properties and their corresponding formation channels. Although results from the Kepler survey indicate that super-Earths are common \citep{howard2012occurrence,fressin2013false}, current surveys have found only three super-Earths (GJ~1214~b, 55~Cnc~e, HD~97658~b) in transit around stars bright enough to enable these planets' detailed atmospheric characterization.  This kind of characterization is crucial for constraining the bulk compositions of these planets, as the presence of a thick atmosphere leads to degeneracies in models of their interior structure \citep[e.g.,][]{rogers2010framework}.

Methods for finding nearby transiting super-Earths include efforts from both the ground and space. Ground-based transit surveys typically focus on observations of smaller M-type stars \citep{berta2012transit,giacobbe2012photometric,kovacs2013sensitivity}, as these have more favorable planet-star radius ratios; however, to date these ground-based surveys have yielded only one super-Earth discovery, that of GJ~1214~b \citep{charbonneau2009super}, and their sensitivity to transits around larger Sun-like stars is limited. Space telescopes offer several advantages over ground-based transit surveys, as they are typically more sensitive and can observe their targets continuously. In 2017, the \textit{TESS} space telescope will begin an all-sky survey of bright, nearby FGKM dwarf stars \citep{ricker2010transiting}. Until that time, searches for transits of super-Earths detected using the radial velocity method provide a promising route to increase the number of such systems. This approach has resulted in the discovery of transits for 55~Cnc~e and HD~97658~b \citep{winn2011super,dragomir2013detects} by the \textit{MOST} space telescope. 

The \textit{Spitzer} space telescope provides a comparable platform for transit surveys of RV-detected super-Earths, and benefits from a higher photometric precision than \textit{MOST}. \cite{gillon2010spitzer,gillon2012improved} have previously utilized \textit{Spitzer} to rule out transits for the super-Earth HD~40307~b and to further characterize the properties of the transiting super-Earth 55~Cnc~e as part of a search for nearby transiting low-mass planets. This paper presents the results of six additional \textit{Spitzer} observations of super-Earth systems. In \S2 we overview the radial velocity data and transit window predictions for these objects. We provide descriptions of the 4.5 $\mu$m \textit{Spitzer} observations along with data reduction methods and transit model analysis in \S3, followed by discussion and conclusions of this work in \S4 and \S5, respectively.

\section{Target System Properties and Radial Velocity Measurements}

\subsection{System Properties}

HD~1461~b has a minimum mass of 8.1$M_{\oplus}$ and orbits a G-type star with a period of 5.77 days. Its eccentricity is estimated to be fairly low at 0.16. Two other planets with minimum masses of 28 and 87$M_{\oplus}$ may exist in the system at periods of 446 and 5017 days but have yet to be confirmed \citep{rivera2010super}. 

HD~7924~b has a minimum mass of 9.26$M_{\oplus}$ and orbits a K-type star with a period of 5.40 days. Eccentricity of the planet is close to zero and fixed at this value in the fits here. No additional planets have been reported in this system \citep{howard2009nasa}.

HD~97658~b has a minimum mass of 8.2$M_{\oplus}$ and orbits a K-type star with a period of 9.50 days. Its eccentricity is estimated at around 0.13. No other planets have been reported in this system \citep{howard2011nasa}. Note that a transit detection and further constraints on planet properties have been recently made by \cite{dragomir2013detects} using \textit{MOST}; see \S4 for details and discussion of this target.

HD~156668~b has a minimum mass of 4.15$M_{\oplus}$ and orbits a K-type star with a period of 4.64 days. Orbital solutions from fits to RV measurements were found for both eccentricities of 0 (fixed) and 0.22, and include the possible effects of one additional planet candidate in the system with a minimum mass of 45$M_{\oplus}$ and a period of 810 days \citep{howard2011nasa2}.

HIP~57274~b has a minimum mass of 11.6$M_{\oplus}$ and orbits a K-type star with a period of 8.14 days. Orbital solutions from fits to RV measurements were found for both eccentricities of 0 (fixed) and 0.20. HIP~57274 also has two additional detected planets in the system, one with a minimum mass of 0.4$M_{Jup}$ and a period of 32 days, and the other with a minimum mass of 0.53$M_{Jup}$ and a period of 432 days \citep{fischer2012m2k}.

GJ~876~d has a minimum mass of 5.85$M_{\oplus}$ and orbits an M-type star with a period of 1.94 days. This planet is estimated to have an eccentricity of about 0.21, and is the inner-most planet in a system with at least three others. These include a second planet with a minimum mass of 0.71$M_{Jup}$ and a period of 30 days, and a third planet with a minimum mass of 2.3$M_{Jup}$ and a period of 61 days \citep{rivera2005similar,correia2010harps}. A fourth planet was also recently detected with a minimum mass of 14.6$M_{\oplus}$ and a period of 124 days \citep{rivera2010lick}.

\subsection{Radial Velocity Ephemerides}

The required length of the observation window, and therefore the constraint that radial velocity measurements placed on ephemerides, limited the initial selection of targets for transit investigation. We chose six targets for this \textit{Spitzer} program that had relatively low uncertainties for their predicted transit times and for most cases required observation windows with durations less than 20 hours. We also excluded any super-Earths with existing \textit{Spitzer} observations spanning predicted transit windows.

Details on the target system properties and the RV determined ephemerides are given in Table 1. We utilize updated ephemerides obtained by a fit to both published and unpublished data for these systems from the California Planet Search group (Howard et al., in prep). Our fits for HD~1461~b appear to prefer an eccentric solution, and we therefore leave eccentricity as a free parameter. For HD~7924~b we assume a circular orbit for the planet, as there was no convincing evidence for a non-zero eccentricity. We used the preliminary transit detection from \cite{henry2011detection} to define our transit window for HD~97658~b; see \S4 for a complete discussion of this target. For HD~156668~b and HIP~57274~b there was marginal evidence for non-zero eccentricities, and we therefore selected modestly longer transit windows spanning both the circular and eccentric predictions for the transit time. The transit times of GJ~876~d are expected to deviate from a linear ephemeris due to perturbations from the other planets in the system, and we therefore calculated individual transit windows spanning the epoch of our observations using an N-body integration of the planet parameters given in Table 2 of \cite{rivera2010lick}.  

\begin{deluxetable*}{lccrrcl}
\tabletypesize{\scriptsize}
\tablecaption{Target System Properties}
\tablecolumns{7}
\tablewidth{0pt}
\tablehead{
\colhead{Target} & \colhead{Stellar Type\tablenotemark{a}} & \colhead{$R_{*}$($R_{\odot})$\tablenotemark{a}} & \colhead{$M\sin i$ $(M_{\oplus})$} & \colhead{Period (days)} & \colhead{\textit{e}} & \colhead{Calculated $T_{0}$\tablenotemark{b}}}
\startdata
HD~1461~b & G3 V & $1.2441\pm0.0305$ & $8.1\pm0.7$\phantom{0} & $5.77267\pm0.00029$ & $0.16\pm0.05$ & $5089.041\pm0.090$\\
HD~7924~b & K0 V & $0.7821\pm0.0258$ & $9.26\pm1.77$ & $5.39699\pm0.00013$ & 0 (fixed) & $5089.757\pm0.037$\\
HD~97658~b & K1 V & $0.68\pm0.02$ & $8.2\pm1.2$\phantom{0} & $9.4957\pm0.0022$\phantom{0} & $0.13\pm0.07$ & $5650.681\pm0.012$\\
HD~156668~b & K3 V & $0.720\pm0.013$ & $4.15\pm0.58$ & $4.64230\pm0.00070$ & 0 (fixed) & \phantom{0}$5855.86\pm0.12$\\
 & & & & $4.64260\pm0.00078$ & $0.22\pm0.08$ & \phantom{0}$5856.18\pm0.23$\\
HIP~57274~b & K5 V & $0.68\pm0.03$ & $11.6\pm1.3$\phantom{0} & $8.1391\pm0.0051$\phantom{0} & 0 (fixed) & \phantom{0}$5932.31\pm0.27$\\
 & & & & $8.1389\pm0.0049$\phantom{0} & $0.20\pm0.10$ & \phantom{0}$5932.25\pm0.32$\\
GJ~876~d & M4 V & $0.3761\pm0.0059$ & $5.85\pm0.39$ & $1.93778\pm0.00002$ & $0.207\pm0.055$ & \phantom{0}$6159.09\pm0.16$\tablenotemark{c}
\enddata
\tablenotetext{a}{Stellar properties for HD~1461, HD~7924, and GJ~876 cited from \cite{vonbraun2013stellar}; other stellar properties cited from RV discovery papers.}
\tablenotetext{b}{JD - 2,450,000}
\tablenotetext{c}{Calculated from an N-body simulation that accounts for perturbations from other planets in the system.}
\end{deluxetable*}

\begin{deluxetable*}{lcccccccc}
\tabletypesize{\scriptsize}
\tablecaption{\textit{Spitzer} Observation Details}
\tablecolumns{9}
\tablewidth{0pt}
\tablehead{
\colhead{Target} & \colhead{UT Start Date} & \colhead{AOR} & \colhead{Duration (hrs)} & \colhead{$n_{img}$\tablenotemark{a}} & \colhead{$t_{int}$ (s)\tablenotemark{b}} & \colhead{$r_{apr}$\tablenotemark{c}} & \colhead{Start - End\tablenotemark{d}} & \colhead{Predicted $T_{c}$\tablenotemark{d}}}
\startdata
HD~1461~b & 2011-08-31 & 42790656 & 12.9 & 355,008 & 0.1 & $2.46$ & $5804.52$ - $5805.06$ & $5804.85\pm0.10$\\
HD~7924~b & 2011-11-01 & 44605184 & 7.9 & 217,600 & 0.1 & $2.37$ & $5866.74$ - $5867.07$ & $5866.92\pm0.04$\\
HD~97658~b & 2012-02-25 & 42608128 & 11.9 & 327,616 & 0.1 & $2.35$ & $5982.73$ - $5983.22$ & $5983.03\pm0.08$\\
HD~156668~b\tablenotemark{e} & 2012-05-03 & 42790912 & 17.3 & 145,856 & 0.4 & $2.86$ & $6050.65$ - $6051.37$ & $6050.84\pm0.12$\\
 & & & & & & & & $6051.17\pm0.23$ \\
HIP~57274~b\tablenotemark{e} & 2012-03-02 & 44273920 & 15.9 & 134,080 & 0.4 & $2.58$ & $5988.93$ - $5989.59$ & $5989.28\pm0.27$\\
 & & & & & & & & $5989.22\pm0.32$ \\
GJ~876~d & 2012-08-19 & 42791424 & 12.4 & 338,560 & 0.1 & $2.63$ & $6158.79$ - $6159.29$ & $6159.09\pm0.16$
\enddata
\tablenotetext{a}{Total number of images.}
\tablenotetext{b}{Image integration time.}
\tablenotetext{c}{Median aperture radius (pixel widths) from noise pixel flux calculation.}
\tablenotetext{d}{JD - 2,450,000}
\tablenotetext{e}{As in Table 1, predicted $T_{c}$ are shown for both zero (first row) and non-zero (second row) eccentricity fits.}
\end{deluxetable*}

\begin{deluxetable*}{lrcccl}
\tabletypesize{\scriptsize}
\tablecaption{Limits on Transit Probability}
\tablecolumns{6}
\tablewidth{0pt}
\tablehead{
\colhead{Target} & \colhead{$M\sin i$ $(M_{\oplus})$} & \colhead{$2\sigma$ limits $(R_{\oplus})$\tablenotemark{a}} & \colhead{Model radii $(R_{\oplus})$\tablenotemark{b}} & \colhead{\textit{a priori}\tablenotemark{c,d}} & \colhead{\textit{a posteriori}\tablenotemark{e}} }
\startdata
HD~1461~b & $8.1\pm0.7$\phantom{0} & $0.64-1.02$ & $1.34$, $1.86$, $2.46$ & $9.1\%$ & $0.15\%$\\
HD~7924~b & $9.26\pm1.77$ & $0.96-1.16$ & $1.38$, $1.93$, $2.54$ & $6.4\%$ & $0.0016\%$\\
HD~97658~b\tablenotemark{f} & $8.2\pm1.2$\phantom{0} & $0.90-1.00$ & $1.34$, $1.87$, $2.47$ & $3.8\%$ & $0.029\%$\\
HD~156668~b\tablenotemark{g} & $4.15\pm0.58$ & $0.82-0.88$ & $1.13$, $1.56$, $2.07$ & $6.7\%$ & $0.44\%$\\
 & & & & & $1.4\%$\\
HIP~57274~b\tablenotemark{g} & $11.6\pm1.3$\phantom{0} & $0.72-0.82$ & $1.45$, $2.03$, $2.69$ & $4.5\%$ & $1.0\%$\\
 & & & & & $1.4\%$\\
GJ~876~d & $5.85\pm0.39$ & $0.52-0.70$ & $1.23$, $1.71$, $2.27$ & $8.3\%$ & $1.1\%$
\enddata
\tablenotetext{a}{Calculated limits on planetary radius derived from fits to light curves using impact parameters of 0 to 0.95.}
\tablenotetext{b}{Model radii derived from the minimum mass found by RV measurements, calculated for planet compositions corresponding to 100\% Fe, 100\% MgSiO$_3$, and 100\% H$_2$O \citep{zeng2013model}.}
\tablenotetext{c}{Transit probability before observations, calculated simply as the ratio of $R_*/a_p$. Non-zero eccentricity will also influence this value; the exact effect is not well-constrained for these targets but for an eccentricity of 0.2 it will lead to at most $\pm0.3-1.0\%$ difference in transit likelihood.}
\tablenotetext{d}{Other factors also influence prior transit likelihood besides geometry; known exoplanet mass occurrence rates combined with minimum mass estimates from RV measurements increase transit likelihood for RV-detected super-Earths \citep[e.g., $P_{tr}=12.5$\% for HD~1461~b,][]{stevens2013transits}.}
\tablenotetext{e}{Transit probability after observations.}
\tablenotetext{f}{See \S4 for discussion of a transit detection outside the observation window \citep{dragomir2013detects}.}
\tablenotetext{g}{Posterior transit probabilities are shown for both zero (first row) and non-zero (second row) eccentricity fits.}
\end{deluxetable*}

\section{\textit{Spitzer} 4.5 micron Data Acquisition and Reduction Methodology}

\subsection{Photometry and Intrapixel Sensitivity}

These observations were obtained using the Infra-Red Array Camera (IRAC) in the 4.5 $\mu$m channel operated in sub-array mode; additional details are shown in Table 2. There is a known instrumental effect during \textit{Spitzer} observations that consists of a ramp up in pixel
sensitivity with time, usually occurring up to an hour in duration at the start of an observation. We therefore padded
our light curves with additional time before the predicted center of transit in case it was necessary to trim
the initial data affected by the ramp, as is standard practice in \textit{Spitzer} analyses. This results in a slightly off-center observation window for each of our targets with regards to their predicted centers of transit.

In all data sets, we extract flux information from the BCD files provided by the \textit{Spitzer} pipeline. We calculate the flux using techniques described in several previous studies \citep{knutson20123,lewis2013orbital,todorov2013warm}. First, we find the center of the stellar point spread function using a flux-weighted centroiding routine, then we perform aperture photometry, testing both fixed and time variable aperture sizes. The fixed aperture radii we tested ranged from 2.0 to 3.0 pixel widths, in steps of 0.1; the time variable apertures were scaled based on the noise pixel parameter \citep{mighell2005stellar}. The noise pixel parameter is proportional to the square of the full width half max of the stellar point spread function, and described by Equation 1 below:

\begin{equation}
\beta=\frac{(\sum \limits_{n}I_{n})^{2}}{\sum \limits_{n}I_{n}^{2}}
\end{equation}

where $I_n$ is the measured intensity of the $n^{th}$ pixel.

We then empirically re-scale the noise pixel aperture radii either as $r = a\sqrt{\beta}$, where $a$ is a scaling factor between 0.8 and 1.7 pixel widths, in steps of 0.1; or alternatively as $r = \sqrt{\beta}+C$, where $C$ is a constant between -0.2 and 1.0 pixel widths, also in steps of 0.1.

We account for variations in intrapixel sensitivity by adopting a nearest
neighbor weighting algorithm, such that
the flux at each time step is normalized by a weighted sum of its
50 nearest neighbors in X and Y space on the pixel array, as described in \cite{knutson20123} and \cite{lewis2013orbital}. 

We then evaluate each of the aperture radius models to find the lowest resulting scatter in the residuals of the fitted light curve. Although the best fit aperture radius varied depending on target, in each case an adjustment based on noise pixel yielded improvements over fixed aperture photometry; however, both methods resulted in null transit detections. The median best fit aperture radius for each light curve is shown in Table 2. Figure 1 shows the raw flux photometry for each observation. Figure 2 shows the corresponding normalized flux photometry after utilizing the nearest neighbor algorithm.

\begin{figure}[h]
\epsscale{1.2}
\plotone{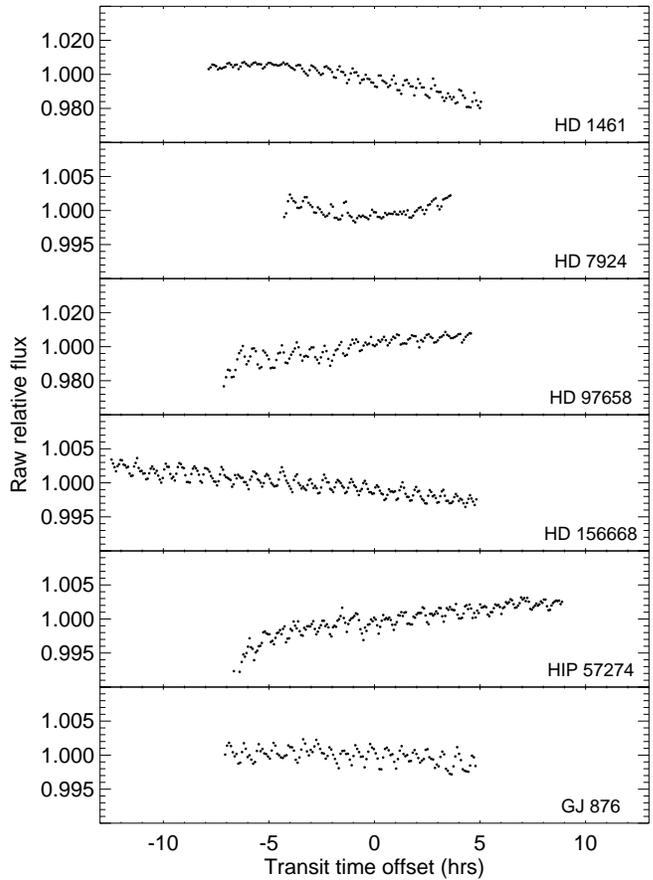}
\caption{Raw \textit{Spitzer} 4.5 $\mu$m light curves. Relative flux is shown binned at 5 minute intervals. Intrapixel sensitivity variations cause distinct sawtooth patterns as a result of the X and Y center position of the stellar point spread function oscillating over time.}
\label{rawfig}
\end{figure}

\begin{figure}[h]
\epsscale{1.2}
\plotone{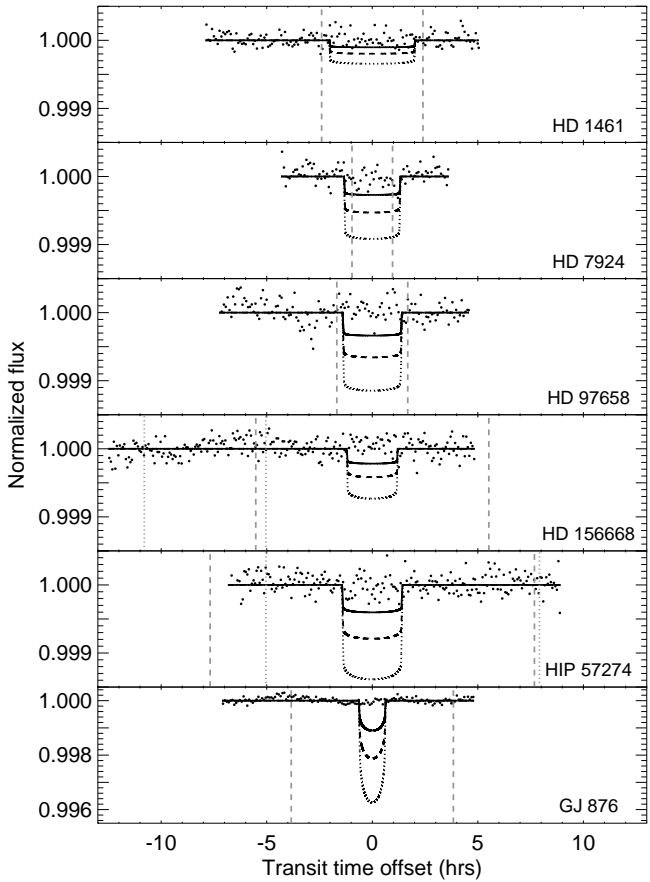}
\caption{Normalized \textit{Spitzer} 4.5 $\mu$m light curves with instrumental effects removed. The normalized flux is shown binned at 5 minute intervals, and transit model light curves for planet radii corresponding to 100\% Fe (black, solid line), 100\% MgSiO$_3$ (black, dashed line), and 100\% H$_2$O (black, dotted line) cases are also shown for comparison. The 1$\sigma$ uncertainties for time of transit center during each light curve are marked as dashed gray vertical lines; for targets HD~156668~b and HIP~57274~b, both uncertainty regions for circular (gray, dotted) and eccentric (gray, dashed) orbital fits are plotted.}
\label{normfig}
\end{figure}

\begin{figure}[h]
\epsscale{1.2}
\plotone{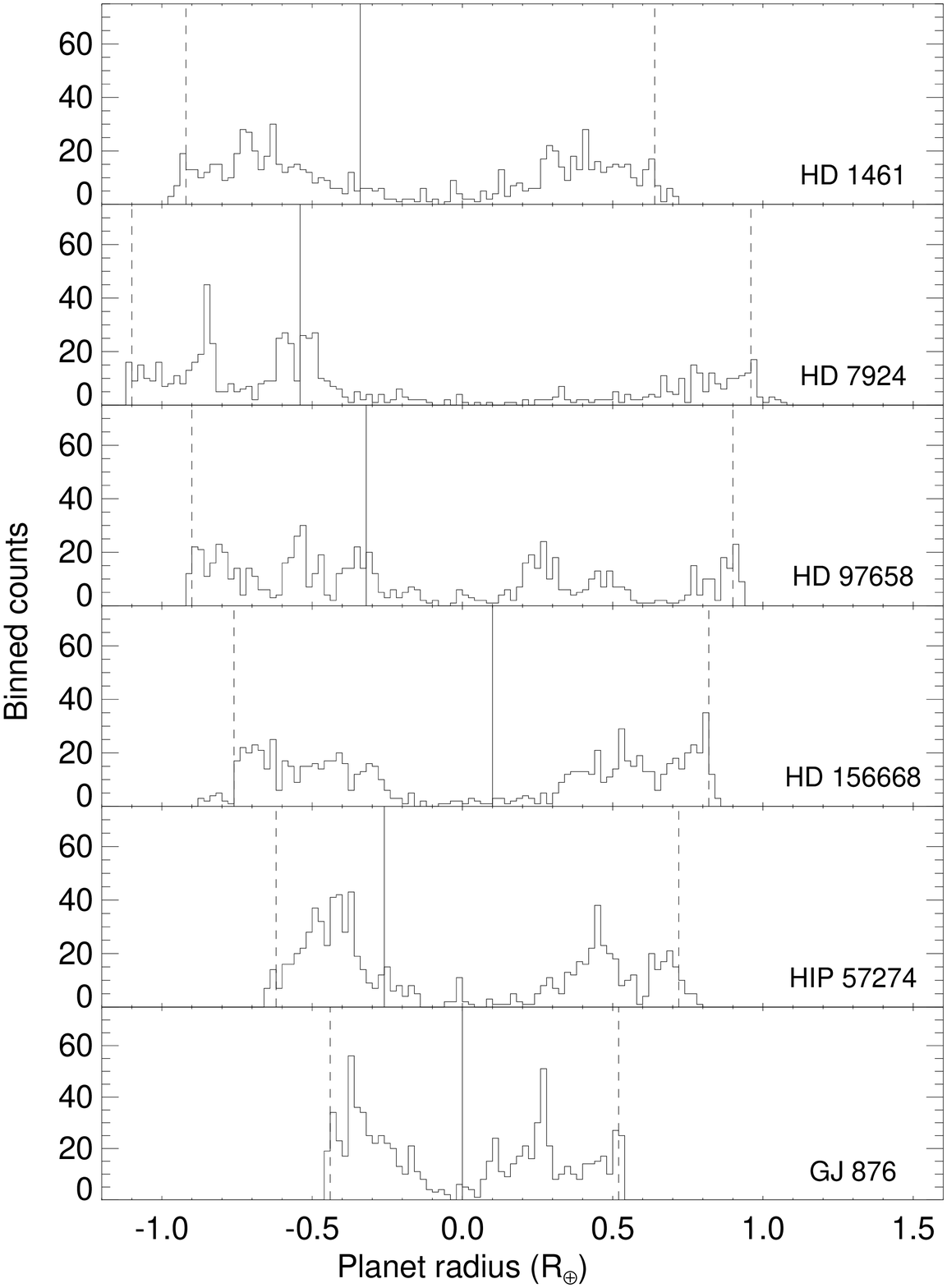}
\caption{Histograms of best fit planetary radius values for a fixed impact parameter of 0. The median best fit radius is given by the solid line, and the 2$\sigma$ uncertainties are given by the dashed lines; these lines may be asymmetric about the median but encompass the 95\% confidence interval. These values characterize the effective noise level of the light curves.}
\label{histfig}
\end{figure}

\begin{figure}[h]
\epsscale{1.2}
\plotone{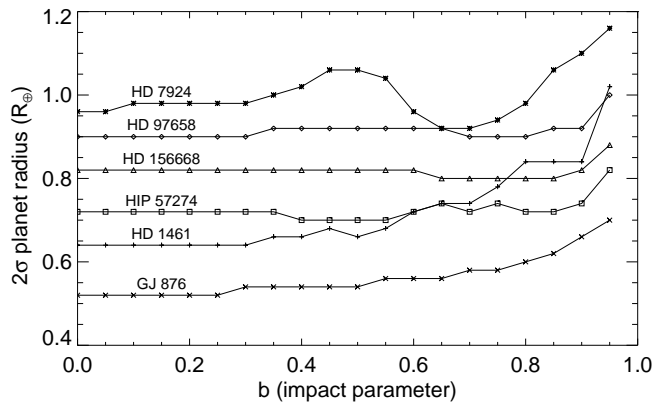}
\caption{Change in detection threshold as a function of fixed impact parameter. For each target, planet radius detection limits are calculated using impact parameters ranging from 0.0 (an equatorial transit) to 0.95 (a grazing transit).}
\label{impactfig}
\end{figure}

\subsection{Transit Models and Uncertainty Estimation}
We fix the orbital parameters for each planet to the values obtained from the radial velocity measurements, and only
the time of transit center, the planet radius, and the impact parameter
are varied in the fits. The forward model for a transit \citep{mandel2002analytic} takes as input these three transit parameters, as well as the orbital period and planet semi-major
axis from RV measurements, and limb darkening coefficients based on
each target's stellar parameters \citep{sing2010limbdark}.

Characterization of transit parameter posterior likelihoods is carried
out using a pseudo-grid search method: given a fixed impact
parameter and transit center time, a best fit planet radius is found
by Levenberg-Marquardt chi-squared minimization. Planet radius is
effectively allowed to be ‘negative’ in these fits by calculating a transit light curve 
using the absolute value of the planet radius, then inverting the curve for negative radius values.
This is done in order not to bias the fits and to better characterize the noise level of the observations.
Figure 3 shows histograms of planetary radii for a fixed impact parameter of zero (an equatorial transit), and fixed transit center times 
that are stepped across the window of observation in increments of approximately 30 seconds. This effectively finds the best fit planet radius at each
location in the light curve. As no significant transits are detected in any of the light curves,
these histograms characterize the magnitude of the combined Gaussian (white) 
and time correlated (red) noise, and therefore provide empirical thresholds for detection of possible transits.
2$\sigma$ limits are calculated that encompass 95\% of the histogram (i.e., 47.5\% of the distribution lies above the median and below the upper limit, 47.5\% below the median and above the lower limit). The 2$\sigma$ limits corresponding to positive planetary radii are then taken
as thresholds for transit detection, as negative radii are non-physical solutions.
These 2$\sigma$ thresholds are shown in Table 3, along with values of planet radii corresponding to models with
100\%~Fe, 100\%~MgSiO$_3$, and 100\%~H$_2$O bulk composition \citep{zeng2013model}, derived using the planet minimum
masses found from RV measurements. Although we expect that a pure iron planet would be very unlikely based on 
current planet formation models, this limiting case allows us to place a strict lower limit on the range of possible radii 
for our target planets. Our estimated radii also assume that the planets have negligible atmospheres, and the 
presence of a thick atmosphere would only serve to increase the transit depth for a given interior composition.

In addition to determining transit detection limits,
we confirm the validity of these limits by inserting artificial transits with depths above
the detection threshold into the data and verifying that we can reliably retrieve them in our fits.
Analysis of these artificially inserted transits yielded consistent
results for detection thresholds of planetary radii. 

In Figure 4 we evaluate the sensitivity of our detection limits to changes in the assumed impact parameter $b$. We find that our limits on planetary radius are fairly insensitive to changes in impact parameter, though this sensitivity varies depending on the target. The limits for HD~97658 and HD~156668 remain nearly constant out to impact parameters of 0.9, while the thresholds for the other targets tend to vary more noticeably but still mostly remain below a planet radius of 1$R_{\oplus}$. The unusual behavior of HD~7924 in this case is likely due to a correlated noise feature in the observed light curve of similar duration and depth as a model transit with an impact parameter of around 0.5 and a planet radius of about 1.1$R_{\oplus}$. The relatively short duration of the HD~7924 light curve influences the sensitivity of this impact parameter test to the noise in the data, but note that even in this case, a planet radius of 1.1$R_{\oplus}$ remains an unphysical solution.

\section{Discussion\label{sec:Discussion}}

As the 2$\sigma$ thresholds for possible transits are in all cases less than the radius of a pure iron core model \citep{zeng2013model}, we therefore conclude that transits for all of our targets are conclusively ruled out within the window of our observations. Table 3 shows the posterior likelihood that the planets may still transit outside the \textit{Spitzer} observation windows. For several cases the probability of transit has been all but eliminated, while for others we calculate the individual probability of transit remains no higher than 1.4\%.

For the case of GJ~876~d, a null transit result is in agreement with the initial photometric measurements of \cite{rivera2005similar}.
However, we note that our non-detection of a transit for HD~97658~b appears on initial inspection to conflict with a recent paper by \cite{dragomir2013detects} announcing the detection of transits with \textit{MOST}.  We centered our \textit{Spitzer} transit window using the predicted transit time from the preliminary ground-based transit detection of \cite{henry2011detection}.  Subsequent follow-up observations by \cite{dragomir2012non} taken within a month of our \textit{Spitzer} observations demonstrated that the planet did not transit at the time predicted by Henry et al.; our data provide additional support for this conclusion.  A later re-analysis by Henry et al. indicated that the apparent transit detection was caused by an airmass effect in the original observations.

Using the updated transit ephemeris from the recent \textit{MOST} detection, we calculate a predicted transit center time of $2455982.17\pm0.06$, approximately 13 hours earlier than the \textit{Spitzer} observation window that started at $2455982.73$. We therefore conclude that our non-detection of a transit for HD~97658~b is consistent with the transit ephemeris reported by \cite{dragomir2013detects}.

\section{Conclusions\label{sec:Conclusions}}

We find no evidence for transits in any of the systems targeted by this survey. There remains some probability that a transit occurred outside the observation window for each target; we know this occurred for HD~97658~b, but the probability is extremely small for our other targets as shown in Table 3. Excluding HD~97658~b, we estimate that the cumulative posterior transit probability for these targets is now only 4.0\%. Their cumulative prior transit probability before observations was 30.5\%; it is therefore not surprising that no transits were detected, but the high value of such transiting systems more than justifies the investment of \textit{Spitzer} time.

Although no transits were detected in this work, future prospects of utilizing this method for super-Earth discovery remain high. By our estimates the majority of stars known to host super-Earths with well-constrained ephemerides have already been observed by either \textit{Spitzer}, \textit{MOST}, or both, but we expect that current and next-generation radial velocity surveys will produce an ever-growing number of such systems in the coming years.  Until the launch of \textit{TESS}, this method remains one of the most promising avenues for detecting transiting super-Earths around bright, nearby stars.

\acknowledgements{
J.-M.D. and N.K.L. acknowledge funding from NASA through the Sagan Exoplanet Fellowship program administered by the NASA Exoplanet Science Institute (NExScI). This work is based on observations made with the \textit{Spitzer Space Telescope}, which is operated by the Jet Propulsion Laboratory, California Institute of Technology, under contract with NASA.
}

\end{document}